\documentclass[11pt]{article}
\usepackage{graphicx}

\usepackage{geometry}
\usepackage{setspace}
\usepackage{graphicx}
\usepackage{amssymb}
\usepackage{epstopdf}
\usepackage{multirow}
\usepackage{extarrows}
\usepackage{rotating}
\usepackage{dsfont}
\usepackage{bm}
\usepackage{mathabx}
\usepackage{endnotes}
\usepackage{tikz}
\newcommand\mathC{\mkern1mu\raise2.2pt\hbox{$\scriptscriptstyle|$}
        {\mkern-7mu\rm C}}              

\newtheorem{definition}{Definition}[section]
\newtheorem{theorem}{Theorem}[section]
\newtheorem{lemma}[theorem]{Lemma}
\newtheorem{proposition}[theorem]{Proposition}
\newtheorem{corollary}[theorem]{Corollary}

\title{Physical Entanglement in Permutation-Invariant Quantum Mechanics} 
           \author{Adam Caulton\\ adam.caulton@gmail.com }
	             
\begin{document}
\maketitle

\begin{abstract}
The purpose of this short article is to build on the work of Ghirardi, Marinatto and Weber (Ghirardi, Marinatto \& Weber 2002; Ghirardi \& Marinatto 2003, 2004, 2005) and Ladyman, Linnebo and Bigaj (2013), in supporting a redefinition of entanglement for ``indistinguishable'' systems, particularly fermions.  According to the proposal, non-separability of the joint state is insufficient for entanglement.  The redefinition is justified by its physical significance, as enshrined in  three biconditionals whose analogues hold of ``distinguishable'' systems.
\end{abstract}

\tableofcontents

\section{Introduction}

In this article, I wish to give support for a redefinition of `entanglement' for ``indistinguishable'' systems; i.e.~systems for which permutation invariance is imposed.  This redefinition was first proposed by  Ghirardi, Marinatto and Weber (Ghirardi, Marinatto \& Weber 2002; Ghirardi \& Marinatto 2003, 2004, 2005), and has recently been endorsed by Ladyman, Linnebo and Bigaj (2013).  My contribution here will be to prove that the proposed redefinition enjoys a physical significance that is not shared by the standard concept, according to which a joint state of a quantum assembly is entangled iff it is non-separable; i.e.~inexpressible as a product state.

The physical significance of the concept, which I here call \emph{GMW-entanglement}, is enshrined in three biconditionals the analogues of which hold for the standard concept of entanglement for ``distinguishable'' systems; i.e.~systems for which permutation invariance is not imposed.  These three biconditionals are:
\begin{enumerate}
\item The joint state of any two-system assembly is entangled iff it violates a Bell inequality.
\item The joint state of any assembly is not entangled iff the constituent systems' states are pure.
\item The joint state of any assembly is not entangled iff the constituent systems' states determine the joint state.
\end{enumerate}
Each of the three biconditionals can be construed in two way: (i) as about the standard notion of entanglement, as applied to ``distinguishable'' quantum systems (i.e.~for which permutation-invariance is not imposed); and (ii) as about GMW-entanglement, as applied to ``indistinguishable'' quantum systems.  The biconditionals under (i) are well-known (the first is a Theorem due to Gisin 1991); under (ii) they are not known.

Proving the first biconditional under construal (ii) is the main work of this paper.  It will be crucial to this proof that a couple of other concepts be understood rather differently, in a permutation-invariant setting, than in the usual setting.  In particular, we need to revise our understanding of what counts as a \emph{local} operation and how to extract the states of constituent systems from the joint state.  The revision of both of these concepts is necessary for the following reason: in the ``distinguishable'' case, these concepts make essential appeal to the factor Hilbert spaces that make up the assembly's joint Hilbert space; and our best understanding of permutation invariance is one in which factor Hilbert space indices---or, equivalently, the order in which they stand in the tensor product---have no physical meaning.  The second and third biconditionals will drop out of a proper revision of these concepts.

In Section \ref{PIinQM}, I briefly review the topic of permutation invariance in quantum mechanics, and argue that its best interpretation is one that treats the invariance as reflecting by a representational redundancy in the standard quantum formalism.  It is the fact of this redundancy which motivates the revisions in the concepts of local operation, constituents' states and entanglement.  In Section \ref{WhatIsE}, Gisin's Theorem and GMW-entanglement are both reviewed, and some confusions cleared up.   Section \ref{BellIn} contains the proposed redefinitions of local operations, constituents' states and entanglement, and proofs of the three biconditionals.

\section{Permutation invariance, symmetric operators and the wedge product}\label{PIinQM}

Permutation-invariant quantum mechanics is standard quantum mechanics with the additional condition of permutation invariance.  We begin with the single-system Hilbert space $\mathcal{H}$ equipped with an algebra of quantities.  From this we define the $N$-fold tensor product $\otimes^N \mathcal{H}$, the \emph{prima facie} state space for $N$ ``indistinguishable'' systems (their indistinguishability is expressed by the fact that any two factor Hilbert spaces are unitarily equivalent).  The joint Hilbert space $\otimes^N \mathcal{H}$ carries a natural unitary representation $U:S_N\to\mathcal{U}(\otimes^N \mathcal{H})$ of the group $S_N$ of permutations on $N$ symbols.  For example, the permutation $(ij)$, which swaps systems $i$ and $j$, is represented by the unitary operator $U(ij)$ defined on basis states (having chosen an orthonormal basis $\{|\phi_k\rangle\}$ on $\mathcal{H}$) by
\begin{eqnarray}
&& U(ij)|\phi_{k_1}\rangle\otimes\ldots\otimes|\phi_{k_i}\rangle\otimes\ldots\otimes|\phi_{k_j}\rangle\otimes\ldots\otimes|\phi_{k_N}\rangle
\nonumber\\
&& \qquad\qquad\qquad\qquad\qquad =
|\phi_{k_1}\rangle\otimes\ldots\otimes|\phi_{k_j}\rangle\otimes\ldots\otimes|\phi_{k_i}\rangle\otimes\ldots\otimes|\phi_{k_N}\rangle
\end{eqnarray}
and then extended by linearity.  

Permutation invariance, otherwise known as the \emph{Indistinguishability Postulate} (Messiah \& Greenberg 1964, French \& Krause 2006), is the condition  on any operator $Q\in\mathcal{B}(\otimes^N \mathcal{H})$, which is to represent a legitimate physical quantity, that it be \emph{symmetric};\footnote{This use of `symmetric' is not to be confused with the condition that $\langle\psi|Q\phi\rangle = \langle Q\psi|\phi\rangle$ for all $|\psi\rangle, |\phi\rangle\in\mbox{dom}(Q)$.} i.e.~for all permutations $\pi\in S_N$ and all states $|\psi\rangle\in\otimes^N \mathcal{H}$,
\begin{equation}
\langle\psi|U^\dag(\pi)QU(\pi)|\psi\rangle = \langle\psi|Q|\psi\rangle
\end{equation}
The representation $U$ is reducible, and decomposes into various  irreducible representations, each irreducible representation corresponding to a different \emph{symmetry type}; namely bosonic states, fermionic states and (if $N\geqslant 3$) a variety of paraparticle states (see e.g.~Tung 1985, Ch.~5). If we consider only the information provided by the symmetric  operators, we treat permutation invariance as a superselection rule, and each  superselection sector corresponds to one of these symmetry types.  

What does it mean to ``impose'' permutation invariance?  Isn't it rather that permutation invariance holds of some operators and not others?  I propose to impose permutation invariance means to lay it down as a necessary condition on any operator's receiving a physical interpretation.  This  justifies, and is justified by, treating the factor Hilbert space labels---i.e.~the order in which single-system operators and states lie in the tensor product---as nothing but an artefact of the mathematical formalism of quantum mechanics.  

What is the justification for interpreting the factor Hilbert space labels in this way?  The ultimate justification  is of course that it leads to an empirically adequate theory.   It is an empirical fact that elementary particles exhibit statistics consistent with their being either bosons or fermions.  But this fact is logically weaker than the claim that factor Hilbert space labels represent nothing.  It \emph{could} be that factor Hilbert space labels represent (or name), for example, the constituent systems, and that the joint state of any assembly of elementary particles remains in the fermionic or bosonic sector under all actual physical evolutions due only to the fact that the corresponding Hamiltonian happens to be permutation-invariant.   Indeed, this interpretative gloss is either explicitly propounded or implicitly assumed by most authors in the literature (e.g.~French \& Redhead 1988; Butterfield 1993; Huggett 1999, 2003; French \& Krause 2006; Muller \& Saunders 2008; Muller \& Seevnick 2009; Caulton 2013).  However, it may be argued that the physical emptiness of the factor Hilbert space labels offers the \emph{best explanation} of the empirical fact that permutation invariance seems always to hold true.  This suggestion is in line with a more general interpretative stance in physics: that any exact symmetry is a symptom of representational redundancy in the corresponding theory's formalism.

The focus of this paper is fermionic states and their compositional structure.  Picking some orthonormal basis $\{|\phi_i\rangle\}$ in $\mathcal{H}$, these states are spanned by states of the form
\begin{equation}
\frac{1}{\sqrt{N!}}\sum_{\pi\in S_N}(-1)^{\deg\pi}|\phi_{i_{\pi(1)}}\rangle\otimes|\phi_{i_{\pi(2)}}\rangle\otimes\ldots \otimes|\phi_{i_{\pi(N)}}\rangle
\end{equation}
and carry the alternating irreducible representation of $S_N$; i.e.~any permutation $\pi$ is represented by multiplication by $(-1)^{\deg\pi}$, where $\deg\pi$ is the \emph{degree} of the permutation $\pi$ (i.e.~the number of pairwise swaps into which  $\pi$ may be decomposed).

 Following  Ladyman, Linnebo and Bigaj (2013), we may use the mathematical apparatus of \emph{Grassmann} or \emph{exterior algebras} to represent fermionic states.  The exterior algebra $\Lambda(V)$ over the vector space $V$ (over the field of complex numbers $\mathbb{C}$)  is obtained by quotienting the tensor algebra $T(V) := \bigoplus_{k=0}^{\infty} T^k(V) = \mathbb{C}\oplus V \oplus (V\otimes V) \oplus (V\otimes V\otimes V) \oplus \ldots$ with the equivalence relation  $\sim$ defined so that $\alpha\sim\beta$ iff $\alpha$ and $\beta$ have the same anti-symmetrization;\footnote{Equivalently, $\Lambda(V)$ is the quotient algebra $T(V)/D(V^2)$ of $T(V)$ by the two-sided ideal $D(V^2)$ generated by all 2-vectors of the form $x\otimes x$. See e.g.~Mac Lane \& Birkoff (1991, \S XVI.6).} i.e.
\begin{equation}
    \Lambda(V) := T(V)/\sim\ .
\end{equation}
For example, $[x\otimes y] = [-y\otimes x]$ and $[x \otimes x] = [\mathbf{0}]$.
We may set $V=\mathcal{H}$, then there is  a natural isomorphism  $\iota$ from the elements of $\Lambda(\mathcal{H})$ onto the vectors of the fermionic Fock space $\mathcal{F_-(H)} := \bigoplus_{N=0}^{\dim\mathcal{H}} \mathcal{A}(\otimes^N\mathcal{H})$. $\iota$ simply takes any $\sim$-equivalence class of degree-$r$ vectors of $T^r(\mathcal{H})$ to the anti-symmetric degree-$r$ vector in $\mathcal{A}(\otimes^r\mathcal{H})$ that is their common anti-symmetrization.  Therefore we may pick out any $N$-fermion state in $\mathcal{A}(\otimes^N\mathcal{H})$ by specifying its pre-image under $\iota$  in $\Lambda^N(\mathcal{H})$ (i.e.~the subalgebra of $\Lambda(\mathcal{H})$ containing only degree-$N$ vectors).

Elements of $\Lambda(V)$ are called \emph{decomposable} iff they are equivalence classes $[x_{i_1}\otimes x_{i_2}\otimes \ldots\otimes x_{i_r}]$ containing product vectors. (Not all elements are decomposable.)
 To anticipate, the decomposable elements of $\Lambda^N(\mathcal{H})$ correspond to states of $\mathcal{A}(\otimes^N\mathcal{H})$ that are non-GMW-entangled.
 
 The product on the exterior algebra is the \emph{exterior} or \emph{wedge product} $\wedge$, defined by its action on decomposable elements as follows: 
  \begin{equation}\label{wedgeprod}
[x_{i_1}\otimes x_{i_2}\otimes \ldots\otimes x_{i_r}] \wedge [x_{i_{r+1}}\otimes x_{i_{r+2}}\otimes \ldots\otimes x_{i_{r+s}}] = [x_{i_1}\otimes x_{i_2}\otimes \ldots\otimes x_{i_{r+s}}] \ ,
 \end{equation}
where  $\{x_1,x_2, \ldots x_{\dim V}\}$ is an orthonormal basis for $V$ and each $i_k\in \{1,2,\ldots, \dim V\}$.  We then extend the definition of $\wedge$ to non-decomposable elements by bilinearity.  (Note that if there is a pair ${i_j} = i_k$ for $j\neq k$, then the righthand side of (\ref{wedgeprod}) is $[\mathbf{0}]$.) For any $\alpha\in\Lambda^r(V)$ and any $\beta\in\Lambda^s(V)$,
 $
 \alpha \wedge \beta = (-1)^{rs}\beta\wedge\alpha \in \Lambda^{r+s}(V)
$.

In the following, I will, like Ladyman, Linnebo and Bigaj (2013), make use of a harmless abuse of notation by referring to anti-symmetric states by their corresponding wedge product.  In particular, given an orthonormal basis $\{|\phi_i\rangle\}$ on $\mathcal{H}$, 
\begin{equation}
|\phi_{i_1}\rangle\wedge|\phi_{i_2}\rangle\wedge\ldots \wedge|\phi_{i_N}\rangle
\end{equation}
will be used as a shorthand for
\begin{equation}
\frac{1}{\sqrt{N!}}\sum_{\pi\in S_N}(-1)^{\deg\pi}|\phi_{i_{\pi(1)}}\rangle\otimes|\phi_{i_{\pi(2)}}\rangle\otimes\ldots \otimes|\phi_{i_{\pi(N)}}\rangle\ .
\end{equation}
It is common to represent an $r$-dimensional subspace of $V$ by a wedge product of $r$ degree-1 vectors (i.e.~vectors in $V$).  Correspondingly,  joint states of $r$ fermions which correspond to decomposable degree-$r$ vectors---i.e.~joint states which are non-GMW-entangled---may be aptly (that is: completely and non-redundantly) represented by  $r$-dimensional subspaces of $\mathcal{H}$.  I return to this point at the end of Section \ref{BellIn}.

\section{What is entanglement?}\label{WhatIsE}

Entanglement is standardly defined formally as the non-separability of the assembly's state; i.e.~a state is entangled iff it cannot be written as a product state (see e.g.~Nielsen \& Chuang 2010, 96).  The physical significance of this definition is underpinned by a biconditional, one half of which is Gisin's Theorem, which applies to assemblies of two (``distinguishable'') subsystems:
\begin{theorem}[Gisin 1991]
Let $|\psi\rangle \in\mathcal{H}_1\otimes\mathcal{H}_2$.  If $|\psi\rangle$ is entangled (i.e.~$|\psi\rangle$ is not a product state), then $|\psi\rangle$ violates a Bell inequality. That is, there is some state $|\chi\rangle \in \mathfrak{h}_1\otimes\mathfrak{h}_2$, where $\mathfrak{h}_1\leqslant \mathcal{H}_1, \mathfrak{h}_2\leqslant \mathcal{H}_2$ and $\dim\mathfrak{h}_1 = \dim\mathfrak{h}_2$, accessible from $|\psi\rangle$ by a local operation, and a triplet of $2\times 2$ matrices $\bm\sigma^{(1)} = (\sigma^{(1)}_x, \sigma^{(1)}_y, \sigma^{(1)}_z)$ on $\mathcal{H}_1$ and a triplet of $2\times 2$ matrices $\bm\sigma^{(2)} = (\sigma^{(2)}_x, \sigma^{(2)}_y, \sigma^{(2)}_z),$ on $\mathcal{H}_2$, each satisfying
\begin{equation}
[\sigma^{(i)}_a,\sigma^{(i)}_b] = 2i\epsilon_{abc}\sigma^{(i)}_c \ ,\quad
\{\sigma^{(i)}_a,\sigma^{(i)}_b\} = 2\delta_{ab}\ ,
\end{equation}
and  four 3-vectors $\mathbf{a}, \mathbf{a}', \mathbf{b}, \mathbf{b}'$ such that
\begin{equation}
\mathcal{I}:=|E(\mathbf{a},\mathbf{b}) - E(\mathbf{a},\mathbf{b}')| + |E(\mathbf{a}',\mathbf{b}) + E(\mathbf{a}',\mathbf{b}')| > 2\ ,
\end{equation}
where\begin{equation}
E(\mathbf{a},\mathbf{b}) := \langle\chi|\mathbf{a}.\bm\sigma^{(1)}\otimes\mathbf{b}.\bm\sigma^{(2)}|\chi\rangle\ ,
\end{equation}
etc.
\end{theorem}
\emph{Proof} See Gisin (1991).
$\Box$

So for the joint Hilbert space $\mathcal{H}_1\otimes\mathcal{H}_2$ to contain any entangled states, we must have $\dim\mathcal{H}_1, \dim\mathcal{H}_2\geqslant 2$.  The other half of the biconditional is the ``easy half'': 
\begin{proposition}
Let $|\psi\rangle \in\mathcal{H}_1\otimes\mathcal{H}_2$.  If $|\psi\rangle$ is not entangled (i.e.~$|\psi\rangle$ is expressible as a product state), then $|\psi\rangle$ satisfies any Bell inequality; that is, for any state $|\chi\rangle$ accessible from $|\psi\rangle$ by a local operation, and any triplet of $2\times 2$ matrices $\bm\sigma^{(1)} = (\sigma^{(1)}_x, \sigma^{(1)}_y, \sigma^{(1)}_z)$ on $\mathcal{H}_1$ and any triplet of $2\times 2$ matrices $\bm\sigma^{(2)} = (\sigma^{(2)}_x, \sigma^{(2)}_y, \sigma^{(2)}_z),$ on $\mathcal{H}_2$, each satisfying
\begin{equation}
[\sigma^{(i)}_a,\sigma^{(i)}_b] = 2i\epsilon_{abc}\sigma^{(i)}_c \ ,\quad
\{\sigma^{(i)}_a,\sigma^{(i)}_b\} = 2\delta_{ab}\ ,
\end{equation}
and any  four 3-vectors $\mathbf{a}, \mathbf{a}', \mathbf{b}, \mathbf{b}'$, then
\begin{equation}
\mathcal{I}:=|E(\mathbf{a},\mathbf{b}) - E(\mathbf{a},\mathbf{b}')| + |E(\mathbf{a}',\mathbf{b}) + E(\mathbf{a}',\mathbf{b}')| \leqslant 2\ .
\end{equation}
\end{proposition}
\emph{Proof.}  Since $|\psi\rangle$ is non-entangled, then it has the form
\begin{equation}
|\psi\rangle = |\phi\rangle\otimes|\theta\rangle
\end{equation} 
for some $|\phi\rangle$ and $|\theta\rangle$.  Any state accessible from $|\psi\rangle$ by a local operation also has this form, so we proceed with $|\chi\rangle=|\psi\rangle$.
Any $E(\mathbf{a}, \mathbf{b})$ then takes the form
\begin{eqnarray}
E(\mathbf{a}, \mathbf{b}) &:=& \langle\psi|\mathbf{a}.\bm\sigma^{(1)}\otimes\mathbf{b}.\bm\sigma^{(2)}|\psi\rangle \\
&=& \langle\phi|\mathbf{a}.\bm\sigma^{(1)}|\phi\rangle\langle\theta|\mathbf{b}.\bm\sigma^{(2)}|\theta\rangle \\
&=:& \alpha\beta
\end{eqnarray}
where $\alpha:=\langle\phi|\mathbf{a}.\bm\sigma^{(1)}|\phi\rangle$ and $\beta:=\langle\theta|\mathbf{b}.\bm\sigma^{(2)}|\theta\rangle$.  If we similarly define $\alpha', \beta'$, then
\begin{eqnarray}
\mathcal{I} &=&|\alpha(\beta - \beta')| + |\alpha'(\beta+\beta')|\ ,
\end{eqnarray}
and since $|\alpha|, |\alpha'|,|\beta|, |\beta'|\leqslant 1$, there is no set of values for which $\mathcal{I}$ exceeds 2.
$\Box$

\begin{corollary}
Let $|\psi\rangle \in\mathcal{H}_1\otimes\mathcal{H}_2$. $|\psi\rangle$ is entangled iff it violates a Bell inequality.
\end{corollary}
This biconditional gives entanglement physical meaning, since the Bell inequalities represent physically realisable results---at least \emph{so long as} we assume that any bounded self-adjoint operator represents a physical quantity.

However, when we turn to permutation-invariant quantum mechanics, the significance of this biconditional should be doubted.  Permutation invariance puts restrictions on the available algebra of quantities for the joint system, and some of those  prohibited quantities are involved in the definition  of the correlation functions $E(\mathbf{a}, \mathbf{b})$.  In a permutation-invariant setting,  $\mathcal{H}_1 = \mathcal{H}_2$ and the only symmetric correlation functions are of the form
\begin{equation}
E(\mathbf{a}, \mathbf{a}) := \langle\psi|\mathbf{a}.\bm\sigma\otimes\mathbf{a}.\bm\sigma|\psi\rangle\ .
\end{equation}
Yet the Bell inequality requires us to independently vary the quantities on each system.  Therefore, under permutation invariance the usual Bell inequality cannot even be constructed.

Two responses are available to us, only one of which is normally taken.  The common response is to refrain from the interpretative strategy endorsed in Section \ref{PIinQM}, and to lift the restriction on the joint algebra placed by permutation invariance.  Permutation-invariance is then construed  as nothing more than a ``dynamical inaccessibility'': the prohibited quantitiess still have physical meaning; it is just that dynamical evolution under them is unavailable to the joint system.  Any proponent of this response may still want to say that the biconditional linking entanglement to the violation of a Bell inequality can be taken seriously, and that therefore non-separability is the right definition of entanglement.

However, as I  argued in Section \ref{PIinQM}, we ought to take a stronger reading of permutation invariance.  Under this reading, any element in the mathematical formalism that is not invariant under arbitrary permutation should not be given a physical interpretation.  In that case, the non-symmetric quantities used in the definition of the correlation functions simply cannot be given any physical meaning.  In that case, we must renounce the idea that non-separability provides a physically adequate definition of entanglement.

These doubts have been expressed by Ghirardi, Marinatto and Weber in a series of papers (Ghirardi, Marinatto \& Weber 2002; Ghirardi \& Marinatto 2003, 2004, 2005).  They propose an alternative definition of entanglement, which have called \emph{GMW-entanglement}.  Although not their explicit definition, it turns out  to be equivalent to following:
\begin{definition}
A joint state is \emph{GMW-entangled} iff it is not the anti-symmetrization of a product state.
\end{definition}
So, for example, the spin-singlet state $|\!\uparrow\rangle\wedge|\!\downarrow\rangle = \frac{1}{\sqrt{2}}(|\!\uparrow\rangle\otimes|\!\downarrow\rangle - |\!\downarrow\rangle\otimes|\!\uparrow\rangle)$ counts as non-GMW-entangled.

It may come as a surprise that a state which we have all learned to think of as \emph{maximally} entangled---indeed, the state most commonly used to illustrate the violation of a Bell inequality---should come out as \emph{non}-entangled on \emph{any} reasonable definition.  But there need be no confusion here. The singlet state is indeed  entangled, \emph{so long as} we have access to the full algebra of bounded operators.  If we do not, as in the case of permutation invariance, then that attribution  needs to be revised.

But aren't \emph{electrons}, which are fermions, and therefore subject to permutation invariance, involved in physical violations of the Bell inequality?  And don't those violations arise in particular when the electrons are prepared in the singlet state?   The answer to both these questions is Yes, but we need to be careful about all of the electrons' degrees of freedom.  As Ghirardi, Marinatto \& Weber (2003, 3) and Ladyman, Linnebo \& Bigaj (2013, 216) point out, the full state in the standard EPRB experiment is
\begin{equation}\label{RealEPR}
\frac{1}{2}\left(|L\rangle_1\otimes|R\rangle_2 + |R\rangle_1\otimes|L\rangle_2\right)\otimes\left(|\!\uparrow\rangle_1\otimes|\!\downarrow\rangle_2 - |\!\downarrow\rangle_1\otimes|\!\uparrow\rangle_2\right)\ ,
\end{equation}
where $|L\rangle$ and $|R\rangle$ represent spatial wavefunctions concentrated at the left-hand and right-hand sides of the lab respectively.  Written using the wedge product, this state is \begin{equation}\label{RealEPR2}
\frac{1}{\sqrt{2}}\left(|L,\uparrow\rangle\wedge|R,\downarrow\rangle - |L,\downarrow\rangle\wedge|R,\uparrow\rangle  \right)\ ,
\end{equation}
which is manifestly \emph{not} expressible as the anti-symmetrization of a product state. So it counts as GMW-entangled. 

But we still lack some way of making sense of Bell inequality violation under permutation invariance---one that agrees with the prevailing belief that state (\ref{RealEPR}) violates a Bell inequality.  It is the purpose of the next Section to do that.

\section{Bell inequalities, local operations and constituent states under permutation invariance}\label{BellIn}

In order to define a Bell inequality in a permutation-invariant setting, we need some way of picking out the subsystems that is permutation-invariant---in particular, we may not appeal to the factor Hilbert space labels.  Our solution, inspired by Ghirardi, Marinatto \& Weber (2002) and Dieks \& Lubberdink (2011), is to appeal to the \emph{states} of the subsystems.  This may be seen as the quantum analogue of Russell's (1905) strategy of picking out objects with a property that is uniquely satisfied.

I illustrate the strategy for the case $N=2$; its generalisation to $N>2$ will be obvious.  The quantum analogue of a 1-place formula is a projector that acts on the single-system Hilbert space $\mathcal{H}$.  So to pick out two subsystems we select two projectors $E_1, E_2$ on $\mathcal{H}$ such that $E_1\perp E_2$, i.e.~$E_1E_2 = E_2E_1 = 0$; I call this condition \emph{orthogonality}.  The orthogonality of the projectors is crucial, since it is necessary and sufficient to ensure that, for any joint state, the two projectors do not select the same subsystem.

However, there is still the danger that one of the projectors, $E_1$ say, will pick out both subsystems.
Since the subsystems are fermions, we can rely on Pauli exclusion to protect us from this if we impose $\dim E_1 = \dim E_2 = 1$.  However, this  condition is far too strong, since it will select subsystems only in the corresponding pure states, and we want to allow the subsystems to occupy mixed states.  (In fact GMW-entangled states are precisely those for which we can ascribe the subsystems mixed states; see Caulton ms.) 

It is sufficient to demand that the joint state $|\psi\rangle$ be an eigenstate of the projector
\begin{equation}\label{IndProj}
E_1\otimes E_2 + E_2\otimes E_1
\end{equation}
with eigenvalue 1; I call this condition \emph{exhaustion}.  Note that this projector is permutation-invariant.  I propose that we interpret it as picking out those joint states in which one subsystem is in a state in ran$(E_1)$ and the other is in a state in ran$(E_2)$.  But it must be emphasized that (\ref{IndProj}) should \emph{not} be interpreted as the quantum disjunction, `Subsystem 1 is in a state in ran$(E_1)$ and subsystem 2 is in a state in ran$(E_2)$ QOR subsystem 1 is in a state in ran$(E_2)$ and subsystem 2 is in a state in ran$(E_1)$.'  The individual disjuncts of this proposition are not permutation-invariant and so have no physical interpretation.  Rather, we must interpret (\ref{IndProj}) \emph{primitively} as the proposition `Exactly one of the subsystems is in a state in ran$(E_1)$ and exactly one of the subsystems is in a state in ran$(E_2)$.'  Interpreting the projector primitively in this way (i.e.~not as a disjunction) is supported by the following fact: if $\dim E_1 = \dim E_2 = 1$, then (\ref{IndProj}) projects onto a single ray in the fermionic Hilbert space, and so could not be a non-trivial disjunction of other propositions.

Therefore our two conditions on what we might call \emph{individuating projectors} $E_1$ and $E_2$ are that they be: (i) orthogonal; and (ii) exhaustive.  A pair of individuating projectors can \emph{always} be found for any given 2-fermion state. (For a proof, see Caulton ms.).  The same is not true for bosonic or paraparticle states.

 Once we have these individuating projectors, we can define operators associated with the corresponding subsystems.  The proposal is simple: any operator $A$ on $\mathcal{H}$ is associated with the subsystem individuated by $E_i$ iff $E_iAE_i = A$.  (Note that if we had demanded that $\dim E_1 = \dim E_2 = 1$, then the algebra of operators associated with each subsystem would be Abelian.)  We can now define permutation-invariant operators on the joint system which act separately on each subsystem; i.e.~they are the permutation-invariant analogues of $A\otimes \mathds{1}$ and $\mathds{1}\otimes B$.  They have the general form
\begin{equation}\label{JointObs}
E_1AE_1\otimes E_2BE_2 + E_2BE_2\otimes E_1AE_1\ , \quad \mbox{where}\ A,B\in\mathcal{B(H)}\ .
\end{equation}

All this leads to the following proposal for what is for a fermionic joint state $|\psi\rangle$ to violate a permutation-invariant Bell inequality:
\begin{definition}
Let $|\psi\rangle \in \mathcal{A(H\otimes H)}$. $|\psi\rangle$ violates a permutation-invariant Bell inequality  iff  there is some state $|\chi\rangle$, accessible from $|\psi\rangle$ by a local operation, and two projectors $E_1, E_2$ on $\mathcal{H}$, such that $E_1\perp E_2$ and 
\begin{equation}
\left(E_1\otimes E_2 + E_2\otimes E_1\right)|\chi\rangle = |\chi\rangle\ ,
\end{equation}
and two triplets of $2\times 2$ matrices $\bm\sigma^{(1)} = (\sigma^{(1)}_x, \sigma^{(1)}_y, \sigma^{(1)}_z), \bm\sigma^{(2)} = (\sigma^{(2)}_x, \sigma^{(2)}_y, \sigma^{(2)}_z)$, satisfying
\begin{equation}
[\sigma^{(i)}_a,\sigma^{(i)}_b] = 2i\epsilon_{abc}\sigma^{(i)}_c \ ,\quad
\{\sigma^{(i)}_a,\sigma^{(i)}_b\} = 2\delta_{ab}\ , \quad 
E_i\sigma^{(i)}_aE_i = \sigma^{(i)}_a\ ,
\end{equation}
and  four 3-vectors $\mathbf{a}, \mathbf{a}', \mathbf{b}, \mathbf{b}'$ such that
\begin{equation}
\mathcal{I}_{PI} := |F(\mathbf{a},\mathbf{b}) -F(\mathbf{a},\mathbf{b}')| + |F(\mathbf{a}', \mathbf{b}) + F(\mathbf{a}',\mathbf{b}')| > 2\ ,
\end{equation}
where\begin{equation}
F(\mathbf{a},\mathbf{b}) := \langle\chi|\left(\mathbf{a}.\bm\sigma^{(1)}\otimes\mathbf{b}.\bm\sigma^{(2)} + \mathbf{b}.\bm\sigma^{(2)}\otimes\mathbf{a}.\bm\sigma^{(1)}\right)|\chi\rangle\ ,
\end{equation}
etc.
\end{definition}
It is important to notice that the formal explication  of a ``local'' operation, used in the definition above, must also change under permutation-invariance.  The guiding physical idea is the same for both: just as, for ``distinguishable'' systems, a local operation is one that acts on each subsystem---i.e.~each factor Hilbert space---independently (and so has product form), so too under permutation-invariance  a local operation should act on each subsystem---as individuated by $E_1$ and $E_2$---independently.  So under permutation invariance a local operation is one whose form is given in (\ref{JointObs}).

We are now ready to prove the biconditional linking GMW-entanglement to the violation of a permutation-invariant Bell inequality.  Each direction of the biconditional will be proved separately.
\begin{theorem}
Let $|\psi\rangle \in\mathcal{A}(\mathcal{H}\otimes\mathcal{H})$.  If $|\psi\rangle$ is not GMW-entangled, (i.e.~$|\psi\rangle$ is  the anti-symmetrization of a product state) then $|\psi\rangle$ satisfies any Bell inequality for symmetric quantities.  That is, for any state $|\chi\rangle$ accessible from $|\psi\rangle$ by a local operation, and any two projectors $E_1, E_2$ on $\mathcal{H}$ such that 
\begin{enumerate}
\item [(i)] $E_1\perp E_2$; and 
\item [(ii)] $
\left(E_1\otimes E_2 + E_2\otimes E_1\right)|\chi\rangle = |\chi\rangle\ ;
$
\end{enumerate}
there is no pair of triplets of $2\times 2$ matrices $\bm\sigma^{(1)} = (\sigma^{(1)}_x, \sigma^{(1)}_y, \sigma^{(1)}_z), \bm\sigma^{(2)} = (\sigma^{(2)}_x, \sigma^{(2)}_y, \sigma^{(2)}_z)$ satisfying
\begin{equation}
[\sigma^{(i)}_a,\sigma^{(i)}_b] = 2i\epsilon_{abc}\sigma^{(i)}_c \ ,\quad
\{\sigma^{(i)}_a,\sigma^{(i)}_b\} = 2\delta_{ab}\ , \quad 
E_i\sigma^{(i)}_aE_i = \sigma^{(i)}_a
\end{equation}
for which, for some choice of four 3-vectors $\mathbf{a}, \mathbf{a}', \mathbf{b}, \mathbf{b}'$,
\begin{equation}
\mathcal{I}_{PI} := |F(\mathbf{a},\mathbf{b}) -F(\mathbf{a},\mathbf{b}')| + |F(\mathbf{a}', \mathbf{b}) + F(\mathbf{a}',\mathbf{b}')| > 2\ ,
\end{equation}
where\begin{equation}
F(\mathbf{a},\mathbf{b}) := \langle\chi|\left(\mathbf{a}.\bm\sigma^{(1)}\otimes\mathbf{b}.\bm\sigma^{(2)} + \mathbf{b}.\bm\sigma^{(2)}\otimes\mathbf{a}.\bm\sigma^{(1)}\right)|\chi\rangle\ ,
\end{equation}
etc.
\end{theorem}
\emph{Proof.}  If $\dim\mathcal{H}<4$, then no pair of triplets of $2\times 2$ matrices satisfying the above conditions can be found. So we assume $\dim\mathcal{H}\geqslant 4$.  Any two projectors $E_1$ and $E_2$ satisfying the above conditions must satisfy $E_1|\phi_1\rangle = |\phi_1\rangle, E_2|\phi_2\rangle = |\phi_2\rangle, E_1|\phi_2\rangle = E_2|\phi_1\rangle = 0$, where $|\psi\rangle$ can be written
\begin{equation}
|\psi\rangle = |\phi_{1}\rangle\wedge|\phi_{2}\rangle \ .
\end{equation}
Any state accessible from $|\psi\rangle$ by a local operation (in the permutation-invariant sense) also has this form, so we proceed with $|\chi\rangle=|\psi\rangle$.
Since $E_1\perp E_2$ and $E_i\sigma^{(i)}_aE_i = \sigma^{(i)}_a$, $\sigma^{(i)}|\phi_j\rangle = 0$ if $i\neq j$ ($i,j\in\{1,2\}$).  Therefore
\begin{eqnarray}
F(\mathbf{a}, \mathbf{b}) &=& 
 \langle\phi_1|\mathbf{a}.\bm\sigma^{(1)}|\phi_1\rangle\langle\phi_2|\mathbf{b}.\bm\sigma^{(2)}|\phi_2\rangle \\
&=:& \alpha\beta
\end{eqnarray}
where $\alpha:=\langle\phi_1|\mathbf{a}.\bm\sigma^{(1)}|\phi_1\rangle$ and $\beta:=\langle\phi_2|\mathbf{b}.\bm\sigma^{(2)}|\phi_2\rangle$.  If we similarly define $\alpha', \beta'$, then
\begin{eqnarray}
\mathcal{I}_{PI} &=&|\alpha(\beta - \beta')| + |\alpha'(\beta+\beta')|\ ,
\end{eqnarray}
and since $|\alpha|, |\alpha'|,|\beta|, |\beta'|\leqslant 1$, there is no set of values for which $\mathcal{I}_{PI}$ exceeds 2.
$\Box$

An important example of a non-GMW-entangled state is 
\begin{equation}
|L,\uparrow\rangle\wedge|R,\downarrow\rangle := \frac{1}{\sqrt{2}}(|L,\uparrow\rangle\otimes|R,\downarrow\rangle +|R,\downarrow\rangle\otimes |L,\uparrow\rangle)\ .
\end{equation}
 No permutation-invariant Bell inequality is violated for this state. 

For the second half of the biconditional, we will need a lemma (also used by  Schliemann \emph{et al} 2001 and Ghirardi \& Marinatto 2004), which is the fermionic analogue of the Schmidt bi-orthogonal decomposition theorem; I merely report it here.
\begin{lemma}\label{antisymmatrix}
For any antisymmetric $d\times d$ complex matrix $A$ (i.e.~$A\in\mathcal{M}(d,\mathbb{C})$ and $A^T=-A$), there exists a unitary transformation $U$ such that $A = UZU^T$, where $Z$ is a block-diagonal matrix of the form
\begin{equation}
Z = \mbox{diag}[Z_1, \ldots Z_r, Z_0], \quad \mbox{where} \ 
Z_i = \left(\begin{array}{cc} 0 & c_i \\ -c_i & 0\end{array}\right) \ \mbox{and}\ c_i \in \mathbb{C}
\end{equation}
and  $Z_0$ is the $(d-2r)\times(d-2r)$ zero matrix.
\end{lemma}
\emph{Proof.}  See Mehta (1989).

\begin{theorem}
Let $|\psi\rangle \in\mathcal{A}(\mathcal{H}\otimes\mathcal{H})$.  If $|\psi\rangle$ is GMW-entangled (i.e.~$|\psi\rangle$ is not the anti-symmetrization of a product state), then $|\psi\rangle$ violates a Bell inequality for symmetric quantities. That is, there is some state $|\chi\rangle$, accessible from $|\psi\rangle$ by a local operation, and two projectors $E_1, E_2$ on $\mathcal{H}$, such that $E_1\perp E_2$ and 
\begin{equation}
\left(E_1\otimes E_2 + E_2\otimes E_1\right)|\chi\rangle = |\chi\rangle\ ,
\end{equation}
and two triplets of $2\times 2$ matrices $\bm\sigma^{(1)} = (\sigma^{(1)}_x, \sigma^{(1)}_y, \sigma^{(1)}_z), \bm\sigma^{(2)} = (\sigma^{(2)}_x, \sigma^{(2)}_y, \sigma^{(2)}_z)$, satisfying
\begin{equation}
[\sigma^{(i)}_a,\sigma^{(i)}_b] = 2i\epsilon_{abc}\sigma^{(i)}_c \ ,\quad
\{\sigma^{(i)}_a,\sigma^{(i)}_b\} = 2\delta_{ab}\ , \quad 
E_i\sigma^{(i)}_aE_i = \sigma^{(i)}_a\ ,
\end{equation}
and  four 3-vectors $\mathbf{a}, \mathbf{a}', \mathbf{b}, \mathbf{b}'$ such that
\begin{equation}
\mathcal{I}_{PI}:=|F(\mathbf{a},\mathbf{b}) -F(\mathbf{a},\mathbf{b}')| + |F(\mathbf{a}', \mathbf{b}) + F(\mathbf{a}',\mathbf{b}')| > 2\ ,
\end{equation}
where\begin{equation}
F(\mathbf{a},\mathbf{b}) := \langle\chi|\left(\mathbf{a}.\bm\sigma^{(1)}\otimes\mathbf{b}.\bm\sigma^{(2)} + \mathbf{b}.\bm\sigma^{(2)}\otimes\mathbf{a}.\bm\sigma^{(1)}\right)|\chi\rangle\ ,
\end{equation}
etc.
\end{theorem}
\emph{Proof.}
$|\psi\rangle$ has the general form
\begin{equation}
|\psi\rangle = \sum_{ij} a_{ij}|\theta_i\rangle\otimes|\theta_j\rangle
\end{equation}
where $a_{ij} = -a_{ji}$.
We can represent $|\psi\rangle$ as a complex $d\times d$ anti-symmetric matrix $A$.  Any unitary transformation $U$ on $\mathcal{H}$ corresponds to the transformation $A\mapsto UAU^T$.  So, given Lemma \ref{antisymmatrix}, we can find a basis $\{|\phi_i\rangle\}$ such  that
\begin{equation}
|\psi\rangle = \sum_{i=1}^{\llcorner \frac{d}{2}\lrcorner}c_i|\phi_{2i-1}\rangle\wedge|\phi_{2i}\rangle \ .
\end{equation}
If $|\psi\rangle$ is GMW-entangled, then we can order the basis vectors so that $c_1,c_2\neq 0$.
 Now define
\begin{equation}
|\chi\rangle := \frac{c_1|\phi_1\rangle\wedge|\phi_2\rangle  + c_2|\phi_3\rangle\wedge|\phi_4\rangle}{\sqrt{|c_1|^2 + |c_2|^2}}\ .
\end{equation}
$|\chi\rangle$ may be obtained from $|\psi\rangle$ by a local, selective operation.
The  idea now is to treat the state $|\chi\rangle$ analogously to the entangled state
\begin{equation}
c_1|\phi_1\rangle\otimes|\phi_2\rangle  + c_2|\phi_3\rangle\otimes|\phi_4\rangle\ ,
\end{equation}
which is subject to Gisin's Theorem.   Define
\begin{equation}
E_1 :=  |\phi_1\rangle\langle\phi_1| + |\phi_3\rangle\langle\phi_3|\ ,
\qquad
E_2 := |\phi_2\rangle\langle\phi_2| + |\phi_4\rangle\langle\phi_4|\ .
\end{equation}
Then it may be checked that $\left(E_1\otimes E_2 + E_2\otimes E_1\right)|\chi\rangle = |\chi\rangle$.   The proof now follows analogously to Gisin (1991).
We  define Pauli-like matrices for the factor spaces.  Let
\begin{eqnarray}
\sigma^{(1)}_x &:=& |\phi_1\rangle\langle\phi_3| + |\phi_3\rangle\langle\phi_1|\\
\sigma^{(1)}_y &:=& -i\left(|\phi_1\rangle\langle\phi_3| - |\phi_3\rangle\langle\phi_1|\right)\\
\sigma^{(1)}_z &:=& |\phi_1\rangle\langle\phi_1| - |\phi_3\rangle\langle\phi_3|
\end{eqnarray}
and
\begin{eqnarray}
\sigma^{(2)}_x &:=& |\phi_2\rangle\langle\phi_4| + |\phi_4\rangle\langle\phi_2|\\
\sigma^{(2)}_y &:=& -i\left(|\phi_2\rangle\langle\phi_4| - |\phi_4\rangle\langle\phi_2|\right)\\
\sigma^{(2)}_z &:=& |\phi_2\rangle\langle\phi_2| - |\phi_4\rangle\langle\phi_4|
\end{eqnarray}
It may be checked that these operators satisfy the conditions above.  Some calculation yields
\begin{eqnarray}
F(\mathbf{a}, \mathbf{b}) &:=& \langle\chi|\left(\mathbf{a.\bm\sigma}^{(1)}\otimes \mathbf{b.\bm\sigma}^{(2)}+\mathbf{b.\bm\sigma}^{(2)}\otimes \mathbf{a.\bm\sigma}^{(1)} \right)|\chi\rangle\nonumber\\
&=& a_zb_z + \frac{2\Re \mbox{e}(c_1c_2^*)}{|c_1|^2 + |c_2|^2}(a_xb_x - a_yb_y) + \frac{2\Im \mbox{m}(c_1c_2^*)}{|c_1|^2 + |c_2|^2}(a_xb_y + a_yb_x)  \\
&=:& a_zb_z + \xi\cos\gamma(a_xb_x - a_yb_y) + \xi\sin\gamma(a_xb_y + a_yb_x)
\end{eqnarray}
where $\xi := \frac{2|c_1c_2|}{|c_1|^2 + |c_2|^2}$ and $\gamma := \arg(c_1c_2^*)$. Note that $0<\xi\leqslant 1$.  We now choose $a_x = \sin\alpha, a_y = 0, a_z = \cos\alpha; b_x = \sin\beta\cos\gamma, b_y = \sin\beta\sin\gamma, b_z = \cos\beta$ to obtain
\begin{equation}
F(\mathbf{a}, \mathbf{b}) = \cos\alpha\cos\beta + \xi\sin\alpha\sin\beta
\end{equation}
Making similar choices for $\mathbf{a}', \mathbf{b}'$, and selecting $\alpha =0, \alpha' =\frac{\pi}{2}$, we obtain
\begin{equation}
\left|F(\mathbf{a}, \mathbf{b}) - F(\mathbf{a}, \mathbf{b}')\right| + |F(\mathbf{a}', \mathbf{b}) + F(\mathbf{a}', \mathbf{b}')| = \left|\cos\beta - \cos\beta'\right| + \xi\left|\sin\beta + \sin\beta'\right|
\end{equation}
We may choose $\cos\beta = -\cos\beta' =: \eta,\ \sin\beta = \sin\beta' = \sqrt{1-\eta^2}$, for which
\begin{equation}
\left|F(\mathbf{a}, \mathbf{b}) - F(\mathbf{a}, \mathbf{b}')\right| + |F(\mathbf{a}', \mathbf{b}) + F(\mathbf{a}', \mathbf{b}')|  = 2(\eta + \xi\sqrt{1-\eta^2}).
\end{equation}
This quantity is maximal for $\eta = \frac{1}{\sqrt{1+4\xi^2}}$, for which it takes the value $\frac{2(1+2\xi^2)}{\sqrt{1+4\xi^2}}$, which is strictly greater than 2 for all $\xi>0$; i.e.~for all non-zero $c_1$ and $c_2$.
$\Box$
\begin{corollary}
Let $|\psi\rangle \in\mathcal{A}(\mathcal{H}\otimes\mathcal{H})$. $|\psi\rangle$ is GMW-entangled iff it violates a permutation-invariant Bell inequality.
\end{corollary}

As mentioned above, an important example of a GMW-entangled state is the EPRB state of two electrons:
\begin{equation}\label{indist}
\frac{1}{\sqrt{2}}\left(|L,\uparrow\rangle\wedge|R,\downarrow\rangle - |L,\downarrow\rangle\wedge|R,\uparrow\rangle  \right)\ .
\end{equation}
If we use the individuating projectors $|L\rangle\langle L|\otimes\mathds{1}_{spin}$ and $|R\rangle\langle R|\otimes\mathds{1}_{spin}$, and the subsystems do not change their location, then this state is physically equivalent (because unitarily equivalent) to the state
\begin{equation}\label{dist}
\frac{1}{\sqrt{2}}\left(|\!\uparrow\rangle_L\otimes|\!\downarrow\rangle_R - |\!\downarrow\rangle_L\otimes|\!\uparrow\rangle_R  \right)\ \in \mathbb{C}^2\otimes\mathbb{C}^2 ,
\end{equation}
in which the subsystems are indexed by their locations $L$ and $R$, and permutation invariance is \emph{not} imposed.\footnote{To be more precise: the Hilbert space $\mathfrak{H}$ spanned by the four fermionic states $\{|L,\uparrow\rangle\wedge|R,\uparrow\rangle, |L,\uparrow\rangle\wedge|R,\downarrow\rangle, |L,\downarrow\rangle\wedge|R,\uparrow\rangle, |L,\downarrow\rangle\wedge|R,\downarrow\rangle\}$, and its associated algebra of bounded operators, is unitarily equivalent to the Hilbert space spanned by the four ``distinguishable''-system states $\{|\!\uparrow\rangle_L\otimes|\!\uparrow\rangle_R, |\!\uparrow\rangle_L\otimes|\!\downarrow\rangle_R, |\!\downarrow\rangle_L\otimes|\!\uparrow\rangle_R, |\!\downarrow\rangle_L\otimes|\!\downarrow\rangle_R\}$, and its associated algebra of bounded operators.  The relevant unitary  is the restriction of $\sqrt{2}|L\rangle_1\langle L|\otimes\mathds{1}^{(1)}_{spin}\otimes |R\rangle_2\langle R|\otimes\mathds{1}^{(2)}_{spin}$ to $\mathfrak{H}$, which sends (\ref{indist}) to (\ref{dist}).  This unitary equivalence is discussed in a more general setting in Caulton (ms.).}  (This is essentially pointed out by Huggett \& Imbo 2009.)  In particular, state (\ref{dist}) violates the standard Bell inequality.

Two more biconditionals characterise entanglement for ``distinguishable'' systems, both of which can be extended to GMW-entanglement under permutation invariance.  The first biconditional is that any joint state $|\psi\rangle$ is not entangled iff the constituent systems occupy pure states.   In a permutation-invariant setting, we may say that constituent systems occupy pure states just in case  individuating projectors $E_1, E_2$ may be found that satisfy our two conditions above (orthogonality and exhaustion) \emph{and} $\dim E_1 = \dim E_2 = 1$.  This conditional is obviously equivalent to $|\psi\rangle$'s being the anti-symmetrization of a product state, i.e.~$|\psi\rangle$'s being non-GMW-entangled.

The second biconditional is that $|\psi\rangle$ is not entangled iff the constituents' states determine the joint state.  (Or in metaphysicians' jargon: $|\psi\rangle$ is not entangled iff the  joint state  supervenes on the constituents' states.)  This biconditional is linked to the first by the following two facts: (i)  the joint state is always pure;  and (ii) pure states are maximally informative (and so \emph{a fortiori} more informative than mixed states).  Since, by the first biconditional, the constituent states are pure iff the joint state is not entangled, the constituents' states carry enough information to collectively determine the joint state iff the joint state is not entangled.  This reasoning carries over for GMW-entanglement in the permutation-invariant setting, so long the symmetry type of the constituents is given.  In particular, for fermions: any collection of $n$ mutually orthgonal single-system pure states serves to determine a unique, non-GMW-entangled joint state: namely, their anti-symmetric combination, or wedge product.

It is important to note that the above suggestions for redefinition entanglement, constituents' states and local operations all rely on our two conditions, (i) orthogonality, $E_1\perp E_2$, and (ii) exhaustion, $(E_1\otimes E_2 + E_2\otimes E_1)|\psi\rangle = |\psi\rangle$, holding; and that (i) and (ii) are impossible to satisfy for certain bosonic states.  (In particular, product states with identical factors: $|\phi\rangle\otimes|\phi\rangle$.)  It is not yet known how to make sense of physical entanglement for such states.

I conclude with a curious feature of fermionic joint states.
It almost goes without saying that in the case of ``distinguishable'' systems \emph{any} given joint state determines the constituents' states; one merely has to perform the appropriate partial trace on the joint system's density operator $|\psi\rangle\langle\psi|$.  One might therefore expect the same to be true under permutation invariance.  However, the partial trace has no physical meaning under permutation-invariance, since it requires selecting a preferred factor Hilbert space,  a selection that, according to our favoured interpretation of permutation invariance, has no physical significance.

We must therefore revise how we extract constituents' states in a permutation-invariant setting.  The most natural alternative---for non-GMW-entangled states at least---is to look to the degree-1 vectors into which the joint state is decomposable.  However, as is familiar from  the use of the wedge product in differential geometry, any degree-$n$ decomposable wedge product (any non-GMW-entangled joint state of fermions) can be decomposed in several---indeed continuum-many---ways, as the anti-symmetric combination of a family of $n$ orthogonal degree-1 vectors ($n$ orthogonal constituents' pure states).  Therefore there is a sense in which a non-GMW-entangled fermionic joint state \emph{fails} to determine the states of the constituents; a phenomenon that is, so to speak, the opposite of  entanglement.

However, any two  families of $n$ degree-1 vectors, into which a non-GMW-entangled fermionic state may be decomposed, have the same linear span. So, as mentioned in Section \ref{PIinQM}, any non-GMW-entangled state of $N$ constituents may be associated with an $N$-dimensional subspace of the single-system Hilbert space $\mathcal{H}$.  We may say that any non-GMW-entangled fermionic joint state fails to determine its constituents' states in a way exactly analogous to the failure of a multi-dimensional vector space  to determine the 1-dimensional rays whose span it is.  This phenomenon is unique to fermions, and suggests one more revision to our standard concepts: the sense in which an assembly is composed from its constituents.  A full discussion of that is a matter for another paper.

\section{References}
\parindent=-12pt

\noindent Bell, J. S. (1964), `On the Einstein-Podolsky-Rosen paradox', \emph{Physics} \textbf{1}, pp.~195-200.

Bell, J. S. (1976), `Einstein-Podolsky-Rosen experiments', \emph{Proceedings of the Symposium on Frontier Problems in High Energy Physics}, Pisa, pp.~33-45.

Butterfield, J.~N. (1993), `Interpretation and identity in quantum theory', \emph{Studies in the History and Philosophy of Science} \textbf{24}, pp. 443-76.

Caulton, A. (2013). `Discerning ``indistinguishable'' quantum systems', \emph{Philosophy of Science}
\textbf{80}, pp.~49-72.

Caulton, A. (ms.). `Individuation, entanglement and composition in permutation-invariant quantum mechanics'. 

Dieks, D. and Lubberdink, A. (2011), `How Classical Particles Emerge from the Quantum World', \emph{Foundations of Physics} \textbf{41}, pp.~1051-1064.

Eckert, K.,  Schliemann, J., Bru\ss, D. and Lewenstein, M. (2002), `Quantum Correlations in Systems of Indistinguishable Particles', \emph{Annals of Physics} \textbf{299}, pp.~88-127.

French, S. and Krause, D. (2006), \emph{Identity in Physics: A Historical, Philosophical and Formal Analysis.}  Oxford: Oxford University Press.

French, S. and Redhead, M. (1988), `Quantum physics and the identity of indiscernibles', \emph{British Journal for the Philosophy of Science} \textbf{39}, pp.~233-46.

Ghirardi, G. and Marinatto, L. (2003), `Entanglement and Properties', \emph{Fortschritte der Physik}
\textbf{51}, pp.~379-387.

Ghirardi, G. and Marinatto, L. (2004), `General Criterion for the Entanglement of Two Indistinguishable Particles', \emph{Physical Review A} \textbf{70}, 012109-1-10.

Ghirardi, G. and Marinatto, L. (2005), `Identical Particles and Entanglement', \emph{Optics and Spectroscopy} \textbf{99}, pp.~386-390.

Ghirardi, G., Marinatto, L. and Weber, T. (2002),  `Entanglement and Properties of Composite Quantum Systems: A Conceptual and Mathematical Analysis', \emph{Journal of Statistical Physics} \textbf{108}, pp.~49-122.

Gisin, N. (1991), `Bell's Inequality Holds for all Non-Product States', \emph{Physics Letters A} \textbf{154}, pp.~201-202.

Greiner, W. and M\"uller, B. (1994), \emph{Quantum Mechanics: Symmetries}, 2nd revised edition.  Berlin: Springer.

Huggett, N. (1999), `On the significance of permutation symmetry', \emph{British Journal for the Philosophy of Science} \textbf{50}, pp.~325-47.

Huggett, N. (2003), `Quarticles and the Identity of Indiscernibles', in K. Brading and E. Castellani (eds.), \emph{Symmetries in Physics: New Reflections}, Cambridge: Cambridge University Press, pp.~239-249.

Huggett, N. and Imbo, T. (2009), `Indistinguishability', in D. Greenberger, K. Hentschel and F. Weinert (eds.), \emph{Compendium of Quantum Physics}, Berlin: Springer, pp.~311-317.

Ladyman, J., Linnebo, \O., and Bigaj, T. (2013), `Entanglement and non-factorizability', \emph{Studies in History and Philosophy of Modern Physics} \textbf{44}, pp.~215-221.

Mac Lane, S. and Birkoff, G. (1991), \emph{Algebra}, Third Edition.  Providence, RI: AMS Chelsea.

 Mehta, M. L. (1989), \emph{Matrix Theory: Selected Topics and Useful Results}.  Delhi: Hindustan
Publishing Corporation.

Messiah, A. M. L. and Greenberg, O. W. (1964), `Symmetrization Postulate and Its Experimental Foundation', \emph{Physical Review} \textbf{136}, pp.~B248-B267.

Muller, F.~A. and Saunders, S. (2008), `Discerning Fermions', \emph{British Journal for the Philosophy of Science}, \textbf{59}, pp.~499-548.

Muller, F. A. and Seevinck, M. (2009), `Discerning Elementary Particles', \emph{Philosophy of Science}, \textbf{76}, pp.~179-200.

Nielsen, M. A. and Chuang, I. L. (2010), \emph{Quantum Computation and Quantum Information}, 10th anniversary edition.  Cambridge: Cambridge University Press.

Russell, B. (1905), `On Denoting', \emph{Mind} \textbf{14}, pp.~479-493.

Schliemann, J., Cirac, J. I., Ku\'s, M., Lewenstein, M. \& Loss, D. (2001), `Quantum correlations in two-fermion systems.' \emph{Physical Review A} \textbf{64}: 022303.

Tung, W.-K. (1985), \emph{Group Theory in Physics.}  River Edge: World Scientific.

\end{document}